\documentclass[fleqn,usenatbib]{mnras}

\usepackage{newtxtext,newtxmath}

\usepackage[T1]{fontenc}

\DeclareRobustCommand{\VAN}[3]{#2}
\let\VANthebibliography\thebibliography
\def\thebibliography{\DeclareRobustCommand{\VAN}[3]{##3}\VANthebibliography}

\usepackage{graphicx}	
\usepackage{amsmath}	
\usepackage{comment}

\DeclareMathOperator{\sech}{sech}

\title[Sgr dSph galaxy dark matter profile]{Nuclear star clusters as probes of dark matter halos: the case of the Sagittarius Dwarf Spheroidal Galaxy}

\author[Herlan et al.]{
Robin Herlan,$^{1}$\thanks{E-mail: robin.herlan@web.de}
Alessandra Mastrobuono-Battisti$^{1, 2, 3}$\thanks{E-mail: alessandra.mastrobuono-battisti@obspm.fr}
and Nadine Neumayer$^{1}$
\\
$^{1}$Max Planck Institut f\"ur Astronomie, K\"onigstuhl 17, 69117 Heidelberg, Germany\\
$^{2}$GEPI, Observatoire de Paris, PSL Research University, CNRS, Place Jules Janssen, 92190 Meudon, France\\
$^{3}$Department of Astronomy and Theoretical Physics, Lund Observatory, Box 43, SE--221 00, Lund, Sweden
}

\date{Accepted XXX. Received YYY; in original form ZZZ}

\pubyear{2015}

\begin{document}
\label{firstpage}
\pagerange{\pageref{firstpage}--\pageref{lastpage}}
\maketitle

\begin{abstract}
The Sagittarius dwarf spheroidal (Sgr dSph) galaxy is currently being accreted and disrupted by the tidal field of the Milky Way.
Recent observations have shown that the central region of the dwarf hosts at least three different stellar populations, ranging from old and metal-poor over intermediate metal-rich to young metal-rich. While the intermediate-age metal-rich population has been identified as part of the galaxy, the oldest and youngest populations belong to M54, the nuclear star cluster (NSC) of the Sgr dSph galaxy.
The old metal-poor component of M54 has been interpreted as at least one decayed GC which was initially orbiting its host galaxy. The youngest population formed in situ from gas accreted into M54 after its arrival at the centre of the host. 
In this work, we use the observed properties of M54 to explore the shape of the inner density profile of the Sgr dSph galaxy. 
To do so, we simulate the decay of M54 towards the centre of the dark matter (DM) halo of its host. We model the DM density profile using different central slopes, and we compare the results of the simulations to the most recent observations of the structural properties of M54. 
From this comparison, we conclude that a GC that decays in a DM halo with a density profile $\propto r^{-\gamma}$ and $\gamma \leq 1$ shows a rotational signal and flattening comparable to those observed for M54. Steeper profiles produce, instead, highly rotating and more flattened NSCs which do not match the properties of M54.
\end{abstract}


\section{Introduction}
The Sagittarius dwarf spheroidal (Sgr dSph) galaxy is one of the closest neighbours and satellites of the Milky Way. 
The spectacular tidal streams that surround the Sgr dSph galaxy are a clear evidence of the ongoing tidal disruption operated by the Milky Way on this galaxy.  
The perihelion of the Sgr dSph galaxy is at about $15$\,kpc from the Galactic centre \citep{2013NewAR..57..100B} and its core is located behind the bulge, a fact that led to its relatively late discovery \citep{i94,i95}.
The apocentre of the leading tail is at a Galactocentric distance of $50$\ kpc, while the apocentre of the trailing tail is between $60$ and $100$\,kpc from the Galactic centre \citep{Belokurov14}.
The current stellar and dynamical masses within the half-mass radius of the Sgr dSph galaxy are $\sim 10^7~{\rm M}_\odot$ and $\sim 10^8~{\rm M}_\odot$, respectively \citep{McConnachie_2012}. 
Before merging with the Milky Way and leaving behind its tidal
streams, the dwarf galaxy had a dark matter (DM) halo mass of up to $10^{11}~M_{\odot}$ \citep{Penarubia10, 2013NewAR..57..100B, Purcell11, Mucciarelli17, Dierickx17} and a stellar mass at least one order of magnitude larger than the current one \citep{2013NewAR..57..100B, Mucciarelli17}. 
\newline
Currently, the Sgr dSph galaxy is devoid of gas and its stellar content is dominated by intermediate-age \citep{2006A&A...446L...1B} and metal-rich stars \citep{2004A&A...414..503B}. \\
There are four globular clusters (GCs) within the main body of the Sgr galaxy:
M54, Terzan 7 as well as Terzan 8 and Arp 2 \citep{1995AJ....109.2533D} and, recently, more candidate GCs have been identified \citep{Bellazzini20,Minniti21,Piatti21}. \\
GCs have often been used to trace the DM distribution of galaxies, through observations, dynamical models and simulations \citep[see e.g.][]{Goerdt06, Lora13, AS16, Alabi16, AS17, Amorisco17, VD18, Meadows20, Leaman20}. 
Part of these studies has been motivated by the existing discrepancy between the density profiles of DM
halos of dwarf galaxies inferred from the observations and the DM density profiles expected according to simulations based on the $\Lambda$-cold DM ($\Lambda$CDM) scenario \citep{2010AdAst2010E...5D}. This discrepancy is known as the ``core-cusp problem''  \citep{Flores94, Moore94, 2010AdAst2010E...5D, Bullock17, Lelli22}.
While observations and dynamical models indicate an
approximately constant DM density \citep{Moore94, Flores94,deblock01a, deblock01b, deblock02, Simon05, Battaglia08, Walter08, 2010AdAst2010E...5D, Oh11, Walker11, Amorisco12, Agnello12, Adams14, Oh15, Relatores19, Relatores19b}, the simulations that most closely resemble what we see
in galaxies today need a steep, $\rho\sim r^{-\gamma}$, power law for the DM density profile \citep{FG84}, with a universal $\gamma$  equal or close to $1$ \citep{NFW96, NFW, Merritt06, Stadel09, Navarro10} or $1.5$ \citep{Moore98,Moore99}, depending on the assumptions and resolution of the simulations. Other studies indicate that the power law shape, although steeper than $\gamma=1$, might not be universal, but rather dependent on the halo mass \citep{Jing00}. 
Possible explanations for the density profile include different baryonic mechanisms that can transform cusps into
cores \citep[see e.g.][]{Yoshida00,Burkert00,Kochanek00,Spergel00,Dave01, Ahn05, Governato10, Inoue11, DC14, Vogelsberger14, Elbert15, Lin16, Bermejo18}, a different DM model \citep{Colin00, Bode01,Lovell12,Marsh15,Maccio15} or observational biases, for example due to the angle of observation \citep{Genina18}. \\
During the last few decades, the technological improvement in observational techniques has
made it possible to resolve the central regions of many galaxies. This led to the discovery
of very compact, massive objects similar to GCs, known as nuclear star clusters \citep[NSCs, for a recent review see, e.g.,][]{Neumayer20}. NSCs can be found in a surprisingly high fraction (50-80 per cent) of all galaxy types \citep{2006ApJ...644L..21F, Sanchez-Janssen2019, Hoyer2021}. With typical half-light radii of just a few pc, 
they are the densest and
most massive stellar clusters in the Universe \citep{Neumayer20}. The mechanisms of formation of NSCs are
currently still under debate: possible scenarios are an in situ formation  where NSC stars form at the galaxy centre from accreted gas \citep{Milosavljevi_2004,Schinnerer_2008,10.1111/j.1365-2966.2009.14954.x,Seth_2008},
and the merger of multiple stellar clusters, where massive star clusters, similar to GCs, inspiral to the galactic centre under the effect of dynamical friction \citep{Tremaine75, bekki_2007,capuzzo2008self,agarwal2011nuclear, 2012ApJ...750..111A, Gnedin14, PMB14, AN15, Arcasedda15, 2017MNRAS.464.3720T, Abbate18}. The actual scenario is likely 
a combination of these channels \citep{Neumayer2011,neumayer_2015,2017MNRAS.464.3720T}. \\
Exactly at the centre of the light distribution of the Sgr dSph galaxy we find its most massive GC, M54, that has properties which suggest that it did not form at the
centre of its host galaxy, but rather migrated there after falling into Sagittarius' potential well
\citep{Monaco05, 2008AJ....136.1147B, Alfaro19, Alfaro_Cuello_2020, Kacharov22}, becoming the NSC of this galaxy. 
At a distance of about $26$\,kpc from the Sun
\citep{2006A&A...447..199R}, M54
is currently, alongside $\omega$ Centauri, a promising contender for the closest known extragalactic NSC.\\
In stark contrast to the other GCs of the Sgr dSph galaxy, M54 is very bright ($M_V=10.0$) and metal-poor \citep{1996AJ....112.1487H,1999AJ....118.1245B}. \\
Using MUSE observations, \cite{Alfaro19} studied the detailed properties of M54, finding that it hosts three distinct stellar populations: an old metal-poor (OMP) population, interpreted as a decayed GC (or the result of the merger of more GCs), a young metal-rich (YMR) population possibly born in situ, and an intermediate-age metal-rich population belonging to the central body of the galaxy. In their follow-up work, \cite{Alfaro_Cuello_2020} investigated the kinematic properties of these populations,
finding that, while the YMR population shows a significant rotation (with amplitude about $5$~km/s) and a high degree of flattening, the OMP population is approximately spherical and shows a low rotational signal (with amplitude $\sim 0.8$~km/s). \\
As the orbital properties and the tidal disruption suffered by a stellar cluster depend on how the mass is distributed in the host galaxy, different DM density profiles imply different final structures and kinematic properties for the NSC of the Sgr dSph galaxy \citep[see][for a study  of the properties of the NSC formed in the Pegasus dwarf galaxy, depending on its density profile]{Leaman20}. 
The aim of this work is to probe the DM content of the Sgr dSph galaxy by simulating the decay of the OMP population of M54 into a cored or cuspy host
galaxy, by means of $N$-body simulations. The results of the simulations can then be compared to observational data \citep[e.g. from][]{Alfaro_Cuello_2020} to determine which is the most probable density profile for the dwarf galaxy's DM halo. \\
In Section \ref{sec:methods} we present our methods and the initial conditions adopted for our simulations. In Section \ref{sect:results} we illustrate our results and in Section \ref{sect:discussion} we discuss those results and draw our conclusions.

\section{Models and simulations}\label{sec:methods}
To explore the mass distribution in the inner regions of the Sgr dSph galaxy, we perform direct $N$-body simulations of the dynamical friction-driven decay and merger of M54 at the centre of its host galaxy. 
We then compare the results of the simulations to recent observations of M54 \citep{Alfaro19, Alfaro_Cuello_2020}.\\
To run the simulations we use phiGRAPE, 
a direct, collisional \textit{N}-body code that uses the 4th order Hermite integration method \citep{10.1111/j.1365-2966.2008.13557.x} adapted to run on graphic processing units \citep{Gaburov2009}. PhiGRAPE is freely available\footnote{See e.g. the version of the code available within the software AMUSE: \url{https://pypi.org/project/amuse-phigrape}}, and widely used to model the formation and evolution of NSCs. 
We assume different mass profiles for both M54 and the Sgr dSph galaxy and rescale the simulation time to the actual evolutionary time as described in the following paragraphs. 

\subsection{Globular cluster initial conditions}\label{sect:ICs}
The current cluster mass of M54 (OMP) is $(1.41 \pm 0.02)\times10^6M_{\odot}$ \citep{2018MNRAS.478.1520B}. This corresponds to the mass of the cluster after the infall to the centre of its host. In this process, the cluster loses part of its stars due to the tidal shredding operated by the galaxy potential.\\
To generate the initial conditions of the infalling cluster, we used the McLuster software \citep{2011MNRAS.417.2300K}.
We modelled the cluster as a \cite{King66} profile with either $W_0=6$ \citep{2005ApJS..161..304M} or $W_0= 8$ \citep{1976ApJS...30..227I, 2008AJ....136.1147B} and half-mass radius $r_h = 6$\,pc \citep{Alfaro19}. The $W_0$ parameter depends on the depth of the central potential well of the cluster, and it is a measure of its concentration.
Our clusters are made of a number $N$ of single mass particles, with $N$ ranging between 6\,000 and 15\,000. We adopted different initial masses for the cluster, starting from a maximum value of $2.0\times10^6M_{\odot}$, down to a minimum value equal to the current mass of M54. 
We put each cluster on a circular orbit and let it decay from an initial galactocentric distance of $50$\,pc\footnote{Using a larger initial distance would result in a significant increase of the computational cost of the simulation, given that a larger portion of the galaxy would need to be modelled as a live system.}. 
 We, therefore, assumed that dynamical friction has circularized the orbits of the clusters prior to the arrival at their starting point set in the simulations. The conditions for circularization are particularly favourable in our cuspy potential, taking also into account that we are simulating a merger between two systems with an extremely low mass-ratio \citep[see][and Sect. \ref{sect:discussion}]{Vasiliev2022}.
\\
Through this set of simulations, we found that an initial mass of $1.5\times10^6M_{\odot}$ best reproduces the observed mass and density profile. Therefore, in this work, we are only presenting simulations done adopting this initial mass. Depending on the initial conditions of our simulations (listed in Table \ref{tab:initial_conditions}), the decay times are $18$, $24$ and $71$~Myr for the cusp, intermediate, and core cases respectively. \\
The choice of using a smaller number of particles ($N$ is either equal to $6\,000$ or $15\,000$) than the actual number of stars is dictated by computational limitations and by the necessity to have the same mass resolution in the cluster and in the DM halo (see Sect. \ref{sect:ICG}). We consider a time re-scaling factor to correct for the effects that this approximation has on the relaxation time of the system (see Sect. \ref{sec:rescaling}). 
To avoid close encounters between the particles, we adopt a softening length of $\varepsilon = 0.01$\,pc. The energy conservation at the end of the simulations is $(E_f-E_i)/E_i\leq5\times10^{-6}$, where $E_i$ and $E_f$ are the initial and final energies, respectively.
\begin{table*}
	\centering
	\caption{Initial conditions for the simulated clusters and the respective Sgr dSph-like galaxies. The name of the simulation, the inner slope of the density profile, $\gamma$, and the mass of each DM particle, $m_{DM}$, are provided for each galaxy. The $W_0$ parameter, the number of particles used to model the cluster, $N_{cl}$, and the mass resolution for the cluster (i.e. the mass of each stellar particle), $m_{cl}$, are provided for each of the simulated systems. The cluster half-mass radius is equal to 6\,pc for all the simulated clusters. The galaxies contain 300,000 particles and have a scaling radius of 6.5 kpc.}
	\label{tab:initial_conditions}
	\begin{tabular}{l|ccc|ccc} 
		\hline
		Name & $\gamma$ & $m_{DM}$ ($M_{\odot}$)& $W_0$ & $N_{cl}$ & $m_{cl}$ ($M_{\odot}$)\\
		\hline
		Core1 & 1/2 & 54 & 6 &  15000 & 100\\
		Core2 & 1/2 & 54 & 8 & 15000 & 100\\
		Inter1 & 1 & 265 & 6 & 6000 & 250\\
		Inter2 & 1 & 265 & 8 & 6000 & 250\\
		Cusp1 & 3/2 & 146 & 6 & 15000 & 100\\
		Cusp2 & 3/2 & 146 & 8 & 15000 & 100\\
		\hline
	\end{tabular}
\end{table*}
\begin{figure*}
	\includegraphics[width=0.8\textwidth]{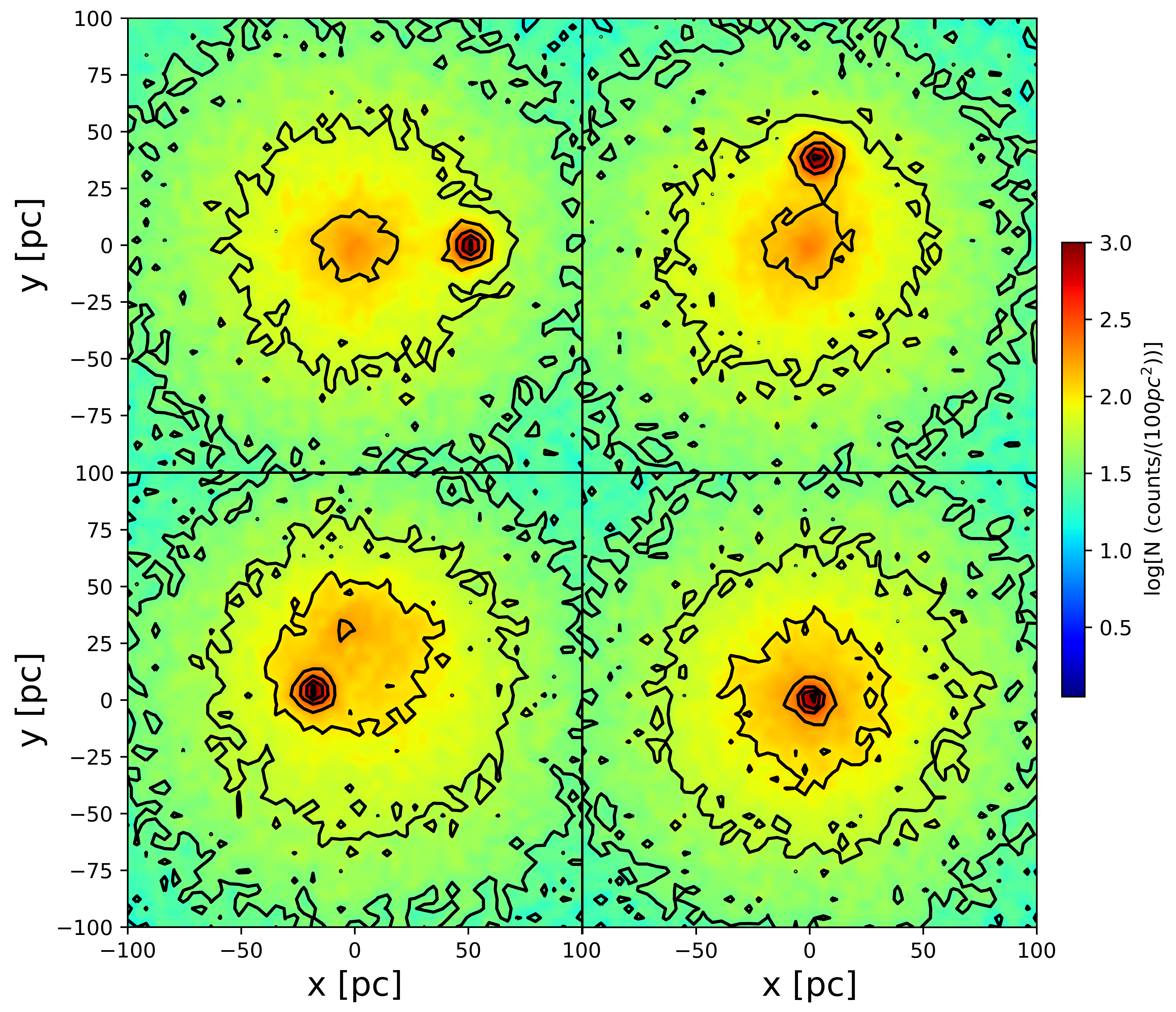}
    \caption{Snapshots of the infall of M54 into the centre of the Sgr dSph galaxy from simulation Inter2. The density maps of the stellar cluster and background DM halo of the Sgr dSph galaxy are shown at four different times. The first snapshot on the top left is taken at the beginning of the simulation, the second one (top right panel) is taken after 3 Myr from the beginning of the decay, the third one (bottom left panel) is taken at  $t=6.5$ Myr and the last one (bottom right panel) is taken at the end of the simulation, that is after 12 Gyr of evolution.  The first three snapshots are in the reference frame of the host galaxy, while the fourth one is centred in the combined centre of density of the galaxy and cluster.}
    \label{fig:InfallI2W6}
\end{figure*}
\subsection{Sgr dSph initial conditions}\label{sect:ICG}
We model the halo of the galaxy as a truncated \cite{1993MNRAS.265..250D} profile 

\begin{equation}
\rho(r) = \frac{(3-\gamma)M}{4\pi}\frac{a}{r^\gamma(r+a)^{4-\gamma}}\frac{2}{\sech\frac{r}{r_{\rm cut}}+\frac{1}{\sech\frac{r}{r_{\rm 
 cut}}}}
\end{equation}
in which $M$ is the total mass of the modelled system, $a$ is the scaling radius, $\gamma$ is a parameter representing the inner slope of the profile, which is restricted to $\gamma = [0,3)$, and $r_{\rm  cut}$ is the truncation radius adopted to reduce the number of particles necessary to represent the galaxy. With the adopted truncation function, we achieve an exponential decrease of the DM density at radii larger than $r_{\rm cut}$, considerably reducing the number of particles necessary to model the system without significantly modifying the density profile in the internal regions.
In our models, we adopt the same initial mass and scale radius assumed by \cite{Purcell11} for their ``heavy Sgr'' model. This model well reproduces the current observed properties of the stream and core of the Sgr dSph galaxy, after merging with the Milky Way. Following this choice, our Sgr dSph has a total mass before truncation equal to $10^{11}\,{\rm M}_\odot$ and a scaling radius $a=6.5$~kpc.  We adopt $r_{\rm cut}=100$~pc, equal to twice the starting distance of the cluster.
While \cite{Purcell11} assumes a NFW profile for the DM halo, we keep the scale-radius and the mass of the halo constant and vary the value of the central slope of the profile. In particular, we adopt a value of $\gamma$ equal to $1/2$, and $3/2$ to represent a galaxy with a shallow cusp, close to a core, and with a central steep cusp as predicted by \cite{Moore99}, respectively. We also explore an intermediate case with $\gamma=1$, equal to the inner slope of the \cite{NFW} profile, in analogy with \cite{Purcell11}.  This is necessary to estimate the effect of different halo profile shapes on the dynamical structure of the NSC and it is a reasonable choice, given the large uncertainties on the initial DM density profile of the Sgr dSph galaxy.
Each halo has been modelled using $N_{DM}=300\,000$ particles. The mass resolution and the total mass of each model depend on the value of the dark matter halo obtained after applying the cut-off, as reported in Table \ref{tab:initial_conditions}. The mass ratio between the dark matter and GC $N$-body particles is kept larger than 0.5 and smaller than 1.5 to avoid spurious effects due to mass segregation. 
The $N$-body models for the galaxy and GC have been generated using the software NEMO \citep{Teuben95}.\\

\subsection{Time rescaling and simulation caveats}\label{sec:rescaling}
As the relaxation strongly affects the final structural properties of the NSC, using a time-consuming collisional approach is necessary to compare the properties of the mock and real M54 cluster. Additionally, the system needs to be simulated as a collisional system because dynamical friction and the dynamical exchange of energy and angular momentum between the cluster and galaxy particles -- that contribute to shape the final NSC properties -- can only be followed using a direct $N$-body code.
 However, due to computational limitations, we model the system with particles more massive than the real stars, and we use a softening length to smooth the two-body interactions. Both these choices affect the relaxation time of the simulated cluster, slowing down the evolution of the system.
To take this effect into account, we rescale the simulation time to the real time  based on the relaxation time of the two systems as described in \cite{10.1046/j.1365-8711.1998.01521.x} and \cite{AMB13}.
Following this approach the real time can be obtained as
\begin{equation}
t = t^*\frac{t_{rx}}{t^*_{rx}}
\end{equation}
where $t^*$ is the simulated time, $t_{rx}$ is the relaxation time of the real system and $t^*_{rx}$
is the relaxation time of the simulated cluster. 
The relaxation time can be written as 
\begin{equation}
t_{rx} = \frac{0.065\sigma^3}{\rho m G^2\ln(\Lambda)}
\label{eq:rel}
\end{equation}
\citep{Spitzer87}, where $m$ is the stellar mass, $\rho$ is the mass density, $\sigma$ is the velocity dispersion of the system and $\Lambda$ is a parameter depending on the maximum and minimum impact distances.
The Coulomb logarithm appearing in Equation (\ref{eq:rel}) can be approximated as $\ln(\Lambda)^*=\ln(r_t/\varepsilon)$ for the simulated system and $\ln(\Lambda)\sim\ln(N)$
for the real cluster, with $r_t$ being the tidal radius of the cluster, $\varepsilon$ the softening length and $N$ the actual number of stars in the cluster. Using the expression for the relaxation time given by \cite{Spitzer87} we can write the real time as
\begin{equation}
t = t^*\frac{m^*\ln{(r_t/\varepsilon)}}{m\ln{(N)}}
\end{equation}
where $m^*$ is the particle mass in the simulation and $m$ is the average stellar mass in the real system (assumed to be equal to $1\,{\rm M}_\odot$)\footnote{We adopted a mean stellar mass equal to $1$\,M$_\odot$. However, we note that the value of the average stellar mass depends on the assumed initial mass function and it is typically lower than $1$\,M$_\odot$. This choice leads to a smaller rescaling factor and, therefore, to a more relaxed final cluster. Our results can be, therefore, considered as lower limits.}.
Since the dynamical friction time is not affected by the mass of the particles, we apply this rescaling only after the cluster has fully decayed to the centre of the galaxy. \\
We neglect the merger of the  Sgr dSph galaxy onto the Milky Way, as this process should not change significantly the properties of the innermost regions of the infalling galaxy. 
Moreover, our simulations do not consider stellar evolution and the presence of binaries in the cluster. These ingredients potentially have a strong effect on the long-term evolution of the cluster, e.g. on the core collapse time and on the cluster dynamical heating due to the presence of the binaries \citep[see e.g.][]{Lynden80, Hills75}. However, the mass loss due to stellar evolution is most relevant during the initial few tens of Myr of the cluster evolution, when massive stars complete their evolution. Afterwards, stellar evolution driven mass loss becomes less impactful on the cluster dynamics and the dynamical mass loss starts to dominate \citep[see e.g.][]{Sollima17}. We, therefore, can consider the initial conditions adopted in this work to represent the cluster after the end  of the initial phase of strong mass loss caused by stellar evolution. Binaries might have a considerable effect on the cluster's long-term evolution and should be included in the calculation. Unfortunately, this is currently impossible due to computational limitations, as we are not yet able to simulate both the cluster and the DM halo on a star-by-star basis. However, we speculate that being more massive than single stars, binaries quickly mass segregate at the cluster centre and mostly affect the evolution of the centremost regions of the NSC, which is the region affected by the largest observational errors.  Additionally, two-body collisional dynamics might not be suited to modelling a DM halo, as the resulting relaxation can cause a spourious evolution in the central dark matter density, which tends to flatten (steepen) cuspy (cored) profiles from the inside-out \citep[see][for a description of this issue in a cosmological context]{Power2003, Diemand2004}. Our DM halo density profiles do not change significantly if simulated in isolation for 12 Gyr, proving that the relaxation has not a significant effect on their mass distribution. The mutual interaction between DM and cluster particles can also contribute to the relaxation. However, the DM density is negligible compared to that of the cluster, minimising the mutual interactions and relaxation effects.  All the limitations listed above will need to be addressed to give a detailed and accurate view of the formation and evolution of M54. 

\subsection{Cluster decay and merger}
Table \ref{tab:initial_conditions} lists the names of the simulations and the relevant parameters describing the cluster and galaxy models.
We simulated the cluster decay towards the centre of the host galaxy and its following disruption and merger.  Before doing so, we checked the stability of the density profile of the galaxy, finding that it remains approximately unchanged over 12\,Gyr if evolved without any decaying cluster.
We consider the cluster as decayed when the Lagrangian radii  containing 20 per cent, 50 per cent and 70 per cent of the total mass of the cluster attain an approximately constant value. The decay time is $18$, $24$ and $71$~Myr for the cusp, intermediate, and core cases respectively.
We observe that the shallower the density profile of the galaxy is,  the longer it takes for the cluster to arrive to the centre of the host galaxy, confirming that dynamical friction is more efficient in cuspy than in core galaxies \citep{Read2006, AnMerr12, AS14}.
The inspiral time does not depend on the internal parameters of the clusters, as dynamical friction is not affected by the cluster mass distribution. 
Figure \ref{fig:InfallI2W6} shows the projected density maps of the stellar and DM particles of the simulated system, for four snapshots of the Inter2 simulation. The cluster, initially visible as an overdensity at $50$\,pc from the centre of the galaxy, follows a decaying orbit that leads it to the centre of the host, where it is visible in the final simulation snapshot (bottom right panel of the figure).  At the end of the simulation, the NSC and the galaxy are co-centered, confirming that the cluster has completed its decay towards the centre of the DM halo. 
After its arrival at the centre of the galaxy, we let the cluster relax until it reaches a final age of 12\,Gyr.   During this interval of time, which is roughly equal to the age of the cluster, collisional effects -- mostly due to two-body interactions between the particles -- drive the relaxation of the system,  leading to its final structure. 

\begin{figure*}
	\includegraphics[width=0.95\textwidth]{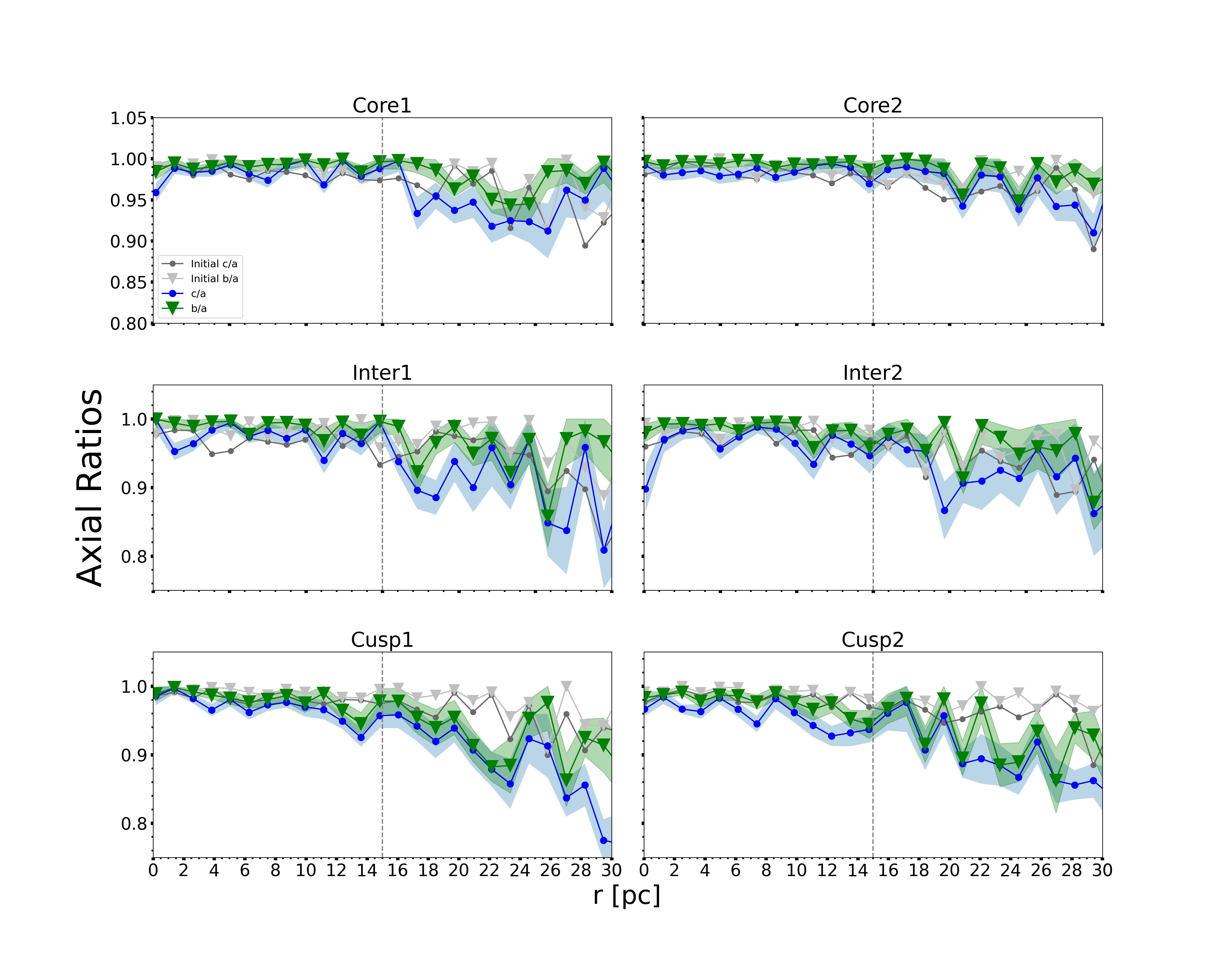} 
    \caption{Axial ratios for all the simulated clusters at the beginning (grey lines) and at the end of the simulation, after 12 Gyrs (green line with triangles for $b/a$ and blue line with bullets for $c/a$). The axial ratios are shown within $30$\,pc from the centre of the cluster, and the name of the simulation is on top of the relative panel.  The shaded area represents the 1$\sigma$ error on the final axial ratio values. The current observational data extend up to $15$\,pc (vertical dashed line).}
    \label{fig:Flatness}
\end{figure*}

\begin{figure*}
\includegraphics[width=0.95\textwidth]{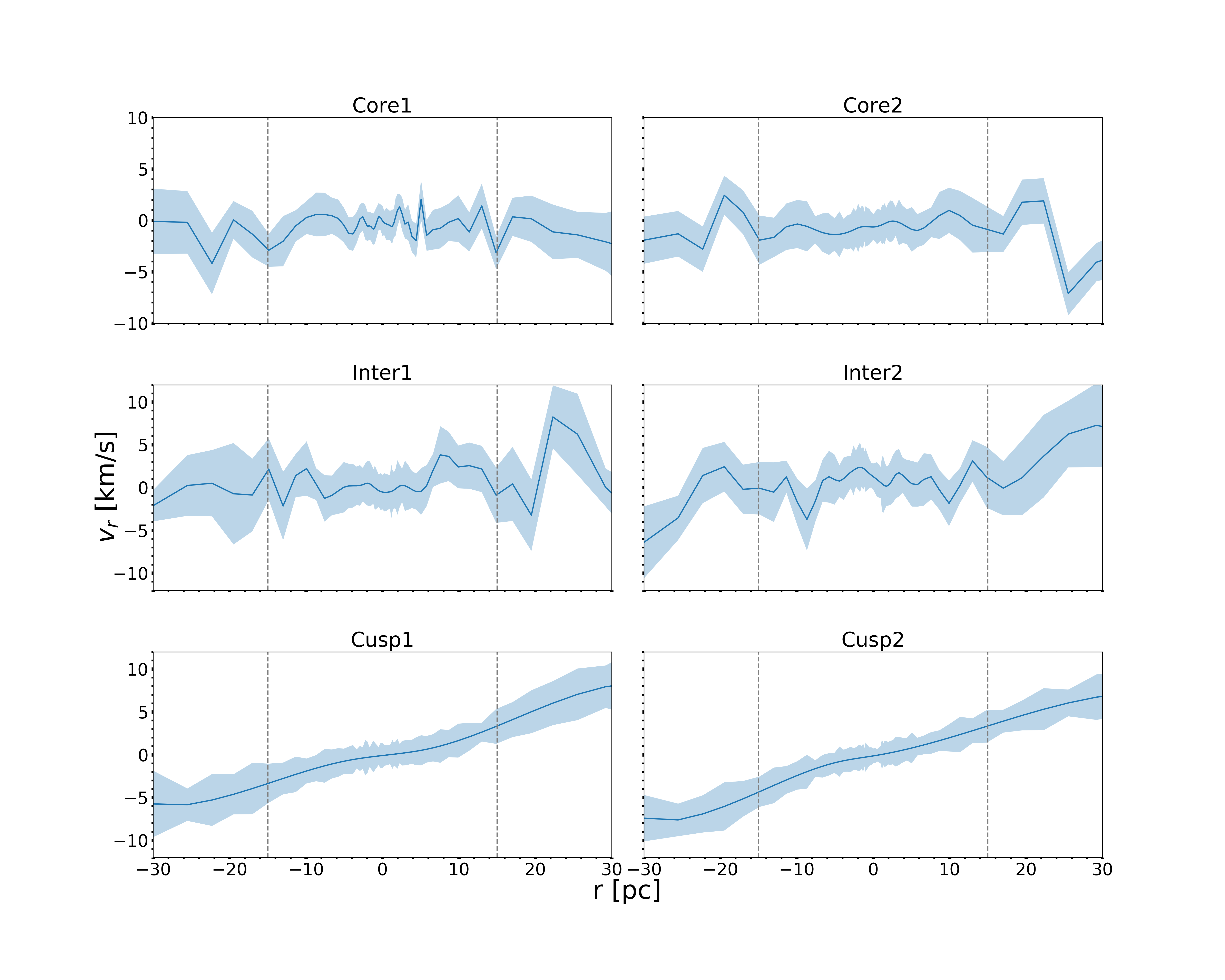} 
\caption{The rotational velocity of M54, $v_r$, within $30$\,pc calculated after 12 Gyr of evolution for each of the simulations, as indicated by the title of each panel. The shaded area displays the $1\sigma$ error on $v_r$. The plot shows a positive correlation between the inner slope of the density profile, $\gamma$, and the velocity curve of M54. While in the core and intermediate cases we find next to no rotation at any distance to the centre, the cusp cases clearly show a rotation of up to 4 km/s at $15$ pc distance from the centre  and of up to $7$ km/s within $30$\,pc.  The shaded area represents the 1$\sigma$ error on the final axial ratio values. The current observational data extend up to $15$\,pc (vertical dashed line).}
\label{fig:Velcurve}
\end{figure*}

\section{Results}\label{sect:results}
We analysed the final structural properties of the decayed cluster to compare them to recent observations of M54 \citep{Alfaro19, Alfaro_Cuello_2020}. 
To allow a better comparison with the observations, we only consider particles bound to the NSC in our analysis. We define such particles as the ones with negative total energy, where the energy is calculated considering the interaction of each particle with all the other particles that were initially in the GC. In this way, we avoid detecting the kinematic signal due to the material tidally stripped from the cluster, which could significantly affect our mock observations.
We focus on the shape and kinematics of the M54-like NSC, as described in the following sections.

\subsection{The morphology of the cluster}
The OMP population observed in M54 shows a low degree of flattening at the half-light radius ($\epsilon_{HL}=0.22$) and in the whole field of view ($\epsilon=0.12$) \citep{Alfaro19}. These values of the ellipticity are corrected for projection effects, taking into account the most probable inclination estimated for M54 (i.e. $60\deg$). 
We quantified the flattening of our final clusters using the inertia tensor as described in \cite{1991ApJ...368..325K} and \cite{2012ApJ...750..111A}. This approach consists in calculating the symmetry axes of the particles enclosed within the ellipsoid $\frac{x^2}{\tau_1^2}+\frac{y^2}{\tau_2^2}+\frac{z^2}{\tau_3^2} = r^2$ as

\begin{equation}
    \tau_1 = \sqrt{\frac{I_{11}}{I_{max}}};
    \tau_2 = \sqrt{\frac{I_{22}}{I_{max}}};
    \tau_3 = \sqrt{\frac{I_{33}}{I_{max}}}
\end{equation}
where $I_{ii}$ are the principal components of the inertia tensor that correspond to the three axis ratios of the system, while $I_{max} = max(I_{11},I_{22},I_{33})$. We identify with $a=I_{max}$ the major axis of the system, while $b$ is the intermediate and $c$ is the minor axis of the ellipsoid. The $c/a$ axial ratio is defined as $c/a=min(I_{11},I_{22},I_{33})/a$ and the intermediate axial ratio $b/a$ corresponds to the intermediate value among the principal components of the inertia tensor.\\
Figure \ref{fig:Flatness} shows the axial ratios obtained for the initial and final simulated systems, along with the $1\,\sigma$ error on the final values\footnote{The errors are calculated as the standard deviation on the axial ratio profile over twenty consecutive snapshots.}. \\
 Within the currently observed range of 15\,pc \citep{Alfaro_Cuello_2020}, Core1 exhibits axial ratios that do not differ from unity, as well as mutually, by more than 4 per cent. Core2 shows axial ratios values which are even closer than for Core1. The deviations from unity in this case are overall smaller than 3 per cent. These two systems are therefore approximately spherical.\\
The intermediate cases exhibit a slightly higher degree of deviation from a spherical shape within $15$\,pc.  In the Inter1 model the deviation from unity of the $c/a$ axial ratio is smaller than 7 per cent at any radius, while the Inter2 model shows a larger flattening (10 per cent) at the centre and becomes flatter (smaller than 5 per cent) outside 2\,pc. 
  A useful parameter to quantify the intrinsic shape of a stellar system is  the triaxiality parameter, defined as $T=(a^2-b^2)/(a^2-c^2)$ \citep[][]{Franx1991}. Oblate and prolate systems have $T = 0$ (same value for the maximum and intermediate axial ratios) and $1$ (same value for the intermediate and minimum axial ratios), respectively, while a triaxial system has $T \sim 0.5$. The NSC central regions (within $10\,$pc) are highly spherical. 
Due to the small number of particles used to model the GCs, the triaxiality parameter shows significant  fluctuations between prolate and oblate values in the region within $20$\,pc, while outside this radius both the Inter1 and Inter2 clusters become oblate. \\
The simulations in which the GC infalls into a cuspy DM halo produce the most asymmetrical M54-like cluster.
Both cluster models show a clear increasing asymmetry with the galactocentric distance, with the $c/a$ ratio reaching differences from unity of about 8 per cent within 15\,pc.   The clusters are approximately spherical within the half-mass radius ($\sim 6\,$pc). According to the triaxiality parameter, the systems are decreasingly triaxial between the half-mass radius and 15\,pc, while outside this radius they become almost perfectly oblate.\\
The core cases show a deviation from sphericity smaller than 10 per cent even within $30$\,pc,  a distance that corresponds to less than half the tidal radius of M54 \citep{1996AJ....112.1487H}. The intermediate and cusp cases show increasing ellipticities at larger distances from the galactic centre. The Inter1 and Cusp1 cases are the most flattened ones, with $c/a$ ratios that can reach a value of $0.8$ within 30\,pc. The Inter2 and Cusp2 cases are slightly less flattened, with $c/a$ larger than $0.85$ at any radius smaller than 30\,pc. \\Currently, it is difficult to observationally constrain the flattening of M54 at large radii, as it is located behind the Galactic Bulge, with a consistent amount of foreground contamination. 
None of our models is able to reproduce the observed ellipticity at the half-light radius, however, at least part of the observed ellipticity could be the result of the relaxation of an initially flattened and rotating YMR population (see Sect. \ref{sect:discussion}). 

\subsection{Kinematical properties of the clusters}
The OMP population in M54 shows a small amount of rotation, with a peak velocity of $1.58\pm1.14$ km/s and an average amplitude of rotation of $0.75\pm 0.55$ km/s \citep{Alfaro_Cuello_2020}.\\
To identify the kinematic major axis of the NSC that forms in our simulations, we projected each system perpendicularly to its maximum angular momentum vector. This is consistent with the most probable inclination angle ($\sim 90^\circ$) estimated by \cite{Alfaro_Cuello_2020}.\\
We evaluated the line-of-sight velocity radial profiles of the cluster at the end of each simulation and in the reference frame of the NSC by considering a mock slit along the identified kinematic major axis of the system. The width of the slit is set to be $5$~pc, a value similar to the MUSE pixel size at the distance of M54. To reduce the noise affecting the results, we averaged over twenty consecutive snapshots to estimate the radial profile of the line-of-sight velocity, calculating the errors as the line-of-sight velocity standard deviation. Figure \ref{fig:Velcurve} shows the velocity curves for our final clusters, plotted within $30$\,pc from the centre of the system. A vertical dashed line shows the extent of the current observational data ($15$\,pc), to allow a direct comparison with \cite{Alfaro_Cuello_2020}.\\
The half-mass radii of the simulated NSCs range between 5.8\,pc and 6.4\,pc; they are, therefore, comparable to the half-mass radius of M54.
The maximum rotational velocity for the simulated clusters is reached at distances larger than 15\,pc.
The NSCs born from the cluster inspiral in a core-like DM halo do not show any significant signs of  rotation within $15$\,pc from the galactic centre. The rotation curve oscillates around zero without any clear coherent trend.\\
The Inter1 case displays a maximum rotational velocity of approximately $2$\,km/s ($2$\,km/s to $5$\,km/s taking into account the errors) within $15$\,pc, however, the rotational trend is not clear. The Inter2 case does not show any clear rotation, as the signal oscillates around $0$\,km/s. In both cases the rotation is smaller than $5$~km/s ($1$\,km/s to $10$\,km/s considering the errors) within $30$\,pc from the centre. The rotation observed within a galactocentric radius of 15\,pc in the core and intermediated cases is lower than the signal detected by \cite{Alfaro_Cuello_2020}. However, as for the ellipticity, part of the observed rotation could come from the relaxation of the embedded rotating YMR population (see Sect. \ref{sect:discussion}). The rotational signal observed at larger distances, which is different for different galaxy profiles, could help to disentangle the cored and intermediate profiles in future observations.
The NSCs that form from a GC that decays in a cuspy galaxy exhibit a high degree of rotation. The Cusp1 case has a maximum velocity of approximately $3$\,km/s ($1$--$5$\,km/s considering the errors) within 15\,pc and of $\sim5$\,km/s ($2$--$10$\,km/s considering the errors) within the central $30$\,pc. The Cusp2 case rotates slightly faster, with a peak velocity of  4\,km/s (3--7\,km/s considering the errors) within 15\,pc and 7\,km/s (3--10\,km/s considering the errors) within $30$\,pc from the centre. These values require an inclination angle close to $0^\circ$ (face-on view) to be  reconcilable with the observations. However, only inclination angles between $60^\circ$ and $90^\circ$ cannot be excluded for M54 \citep{Alfaro_Cuello_2020}. Inclinations within this allowed range are not sufficient to reduce the observed signal to the observed level. \\

Figure \ref{fig:VelmapC2W8} in Appendix \ref{app:vel_maps} shows the velocity maps the NSCs produced in our simulations. The maps are built considering an edge-on view on the systems  and are an additional result that can be compared with observations, e.g. done with integral field unit instruments, like MUSE. 
The core and intermediate cases show no clear signs of rotation, confirming what is seen for their velocity curves.
The substructures observed in the velocity maps are caused by the smoothing procedure and are not physical effects.
The cusp cases show clear signs of rotation, in agreement with what is found for their rotation curves. This rotation is a signature of the merger. Since we are considering only bound particles, the signal is not due to the tidally stripped material, but only to the orbital angular momentum retained by the cluster particles and translated into the NSC rotation.
This result suggests that the faster decay happening in cuspy DM halos\footnote{Same radius circular orbits have larger orbital angular momentum in cuspy galaxies, due to the higher enclosed mass.}, where  dynamical friction is more efficient and cluster particles have less time to interact with the field redistributing their energy and angular momentum, the orbital angular momentum is retained more efficiently leading to a final higher NSC rotation. 
 \\

\section{Discussion and conclusions} \label{sect:discussion}
The properties of an NSC formed through the infall and merger of one or more GCs strongly depend on the mass distribution at the centre of the host galaxy \citep{Leaman20}. Observations and simulations are in apparent conflict with the prediction of the shape of DM halos in galaxies. This ``core-cusp'' problem is still an open question \citep[see e.g.][]{FG84, 2010AdAst2010E...5D, Massari2018}. We used the observed properties of M54 to infer the shape of the DM distribution in the parent galaxy of this system, the Sgr dSph galaxy.\\
We simulated the infall of an M54-like cluster towards the centre of three galaxies with different DM density profiles: a core-like DM halo represented with a \cite{1993MNRAS.265..250D} density profile with central slope equal to $\gamma=1/2$, an intermediate case with $\gamma=1$ and a cusp-like DM halo with $\gamma=3/2$. We used two different models for M54, with distinct central concentrations. The results seem not to strongly depend on this parameter. \\ 
Our clusters start to decay from an initial galactocentric distance of $50$~pc. This choice implies that dynamical friction acted on them in the past, bringing the clusters from their formation orbit to the initial distance assumed in our simulations. This is a delicate assumption, especially in the case of $\gamma=0.5$, as dynamical friction is known to be strongly suppressed in galaxies hosting a core-like DM halo. Dynamical friction, indeed, gets to a core-stalling phase at a distance where the enclosed galaxy mass is approximately equal to the cluster mass\footnote{This is expected to happen at a distance of $90$~pc in the core case, of $20$~pc in the intermediate case and of $3$~pc in the cusp case. We do not observe any core stalling in these two latter cases.} \citep[see e.g.][]{Goerdt10, Arca14, Petts16}. 
 \\
However, the core-stalling can be reduced or even avoided in triaxial galaxies and for clusters on elliptical orbits \citep{Chandrasekhar43, Binney77, Pesce92}. These two conditions are often fulfilled in real galaxies. Therefore, if M54 was born in a galaxy with a core-like DM halo, it could have overcome the stalling phase arriving at the centre of its host, where it is observed today.  Our reference observations are described in \cite{Alfaro19} and \cite{Alfaro_Cuello_2020}. In these papers, the authors accurately evaluate the membership probability based on the line of sight velocity and velocity dispersion of the stars in their field of view and consider only stars with membership probability larger than 70 per cent in their analysis.  Most of the observed stars have membership probabilities close to unity, and therefore the observational result should not be affected by the detection of escaped stars. With these data, they find an average rotational amplitude of $(0.75\pm 0.55)$ km/s  and a  maximum rotation of $(1.58\pm1.14)$\,km/s for the OMP population observed in M54. This population is identified with the one, or more, decayed GCs that formed Sagittarius' nucleus. The OMP population is flattened with an ellipticity at the half-light radius $\epsilon_{HL}=0.22$. The ellipticity in the whole field of view is equal to $\epsilon=0.12$ \citep{Alfaro19}. More recently, \cite{Kacharov22} used multicomponent dynamical models to constrain the rotation and flattening of the different populations in M54. With this approach, they confirmed the observational results, finding that, while the YMP population is for sure flattened and significantly rotating, the OMP population shows a slower rotation and has a shape that is consistent with being spherical within one sigma.
We find that the clusters that decay in a cuspy DM halo acquire a significant amount of rotation once settled at the galactic centre. The NSC that forms in these cases is  flattened at any distance from the centre.  The system is  almost spherical within its half-mass radius, then it shows a decreasing triaxiality within 15\,pc and it becomes oblate outside this radius. The flattening increases with the galactocentric distance, reaching axial ratio values of about $ 0.80$ at $30$~pc from the centre of the galaxy. The morphology and kinematics of the NSC are possibly due to the higher orbital angular momentum of the GC in such a system.  We note that the initial orbital angular momentum of the cluster is about 30 per cent higher in the cusp cases than in the core cases.  While initially low angular momentum (i.e. highly eccentric) orbits could lead to a lower amount of rotation, dynamical friction is expected to efficiently circularize the orbits of low mass-ratio satellites decaying in cuspy potentials \citep{Vasiliev2022}, justifying our choice for an initially high orbital angular momentum. In addition, in a cuspy profile the cluster decays more quickly compared to what happens in shallower potentials. This decreases the possibility for the cluster to interact with the field stars and to redistribute its angular momentum among the DM particles. As the time available to relax is approximately the same in each modelled case, the high initial angular momentum and fast decay are the most important factors in determining the final NSC properties. \\
In the intermediate cases, once it becomes an NSC, the decayed GC shows a mild to unclear rotational signal compared to the NSC formed in the cuspy DM halo. The velocity curve is approximately flat within $15$~pc from the centre of the galaxy. The rotational velocity increases outside this radius and reaches values up to $5$--$10$km/s within $30$\,pc from the galactic centre. The clusters that form in the intermediate case are almost spherical within 10\,pc. These clusters show a triaxiality parameter that oscillates between oblate and prolate within 20\,pc, to then become oblate outside this radius. However, the flattening is  smaller than in the cusp cases. \\
In the cored DM density profiles, the newly formed NSC shows no clear signs of rotation. This lack of a rotational signal is reflected in the approximately spherical shape of the system.\\
A comparison of the simulation results to the observations leads to the conclusion that a cuspy DM density profile would have led to rotational velocities higher than what is observed for M54. 
A cored or an intermediate profile, however, results in no significant rotation of M54  even assuming the highest possible initial orbital angular momentum (i.e. a circular orbit) for the decaying clusters, implying that neither of the profiles can fully explain the small amount of observed rotation. Nevertheless, the core and intermediate profiles, which might show a small degree of rotation, are much closer to the observed properties than the cusp. 
 \\
As argued by \cite{Alfaro_Cuello_2020} and \cite{Kacharov22}, at least part of the rotation and flattening observed for the OMP population of M54 could be the result of the angular momentum transferred from the faster rotating and more elliptical YMR population to the OMP during the relaxation process of the stars belonging to the younger component. 
It is interesting to note that the flattening and rotation might also strongly depend on the strength of the tidal field in which the cluster is embedded \citep[][]{Goodwin97}. Stars with high angular momentum are quickly stripped from the cluster in a cuspy tidal field, leading to a more spherical shape \citep[][]{Longaretti96}. On the contrary, if the cluster is in a core-like tidal field, those stars will be retained, leading to a long-standing flattened and rotating structure. These aspects,  which have been partially investigated in \cite{Alfaro_Cuello_2020} through $N$-body simulations, will be further explored in a future work in which we will take into consideration a flattened and rotating YMR population in an M54-like cluster decaying in differently shaped DM halos.
\\ 
In conclusion, results  suggest that an initial DM density profile with $\gamma \leq 1$ provides the best fit to the observational properties of M54.\\
Moreover, we find that the dynamical properties of M54, or more in general of NSCs in dwarf galaxies, can be used to investigate the DM density profile of the host galaxy. 
In the case of the Sgr dSph galaxy, a larger number of models exploring a wider parameter space, both for the cluster and for the galaxy, and taking into consideration the stellar component of the Sgr dSph galaxy as well as the ongoing merger between the Sgr dSph galaxy and the Milky Way will be necessary to allow a more detailed comparison with the observations and further constrain the shape of the central regions of the galaxy. Observations spanning a larger radial range will be useful to estimate the inner slope of the DM density profile of the Sgr dSph galaxy.
    
\section*{Acknowledgements}
 The authors thank Anna Sippel, Ralf Klessen and Andrew Gould for their valuable comments, as well as Manuel Arca Sedda for providing them with the dark matter halo models used in some preliminary simulations. We thank the referee for the helpful comments that improved the paper. RH, AMB and NN acknowledge support by Sonderforschungsbereich SFB 881 ``The MilkyWay System'' (subproject A7, A8 and B8) -- Project-ID 138713538 -- of the German Research Foundation (DFG).  AMB and NN acknowledge support from DAAD PPP project number 57316058 ``Finding and exploiting accreted star clusters in the Milky Way''. AMB acknowledges funding from the European Union\textquotesingle s Horizon 2020 research and innovation programme under the Marie Sk\l{}odowska-Curie grant agreement No 895174.
 
\section*{Data Availability}
The outputs of the simulations and data analysis scripts are available upon request.


\bibliography{bibfile.bib}{}
\bibliographystyle{mnras}




\appendix




\section{Velocity maps of the decayed globular clusters}\label{app:vel_maps}

\begin{figure*}
	\includegraphics[trim={2.3cm 1 2.3 1},clip, width=0.45\textwidth]{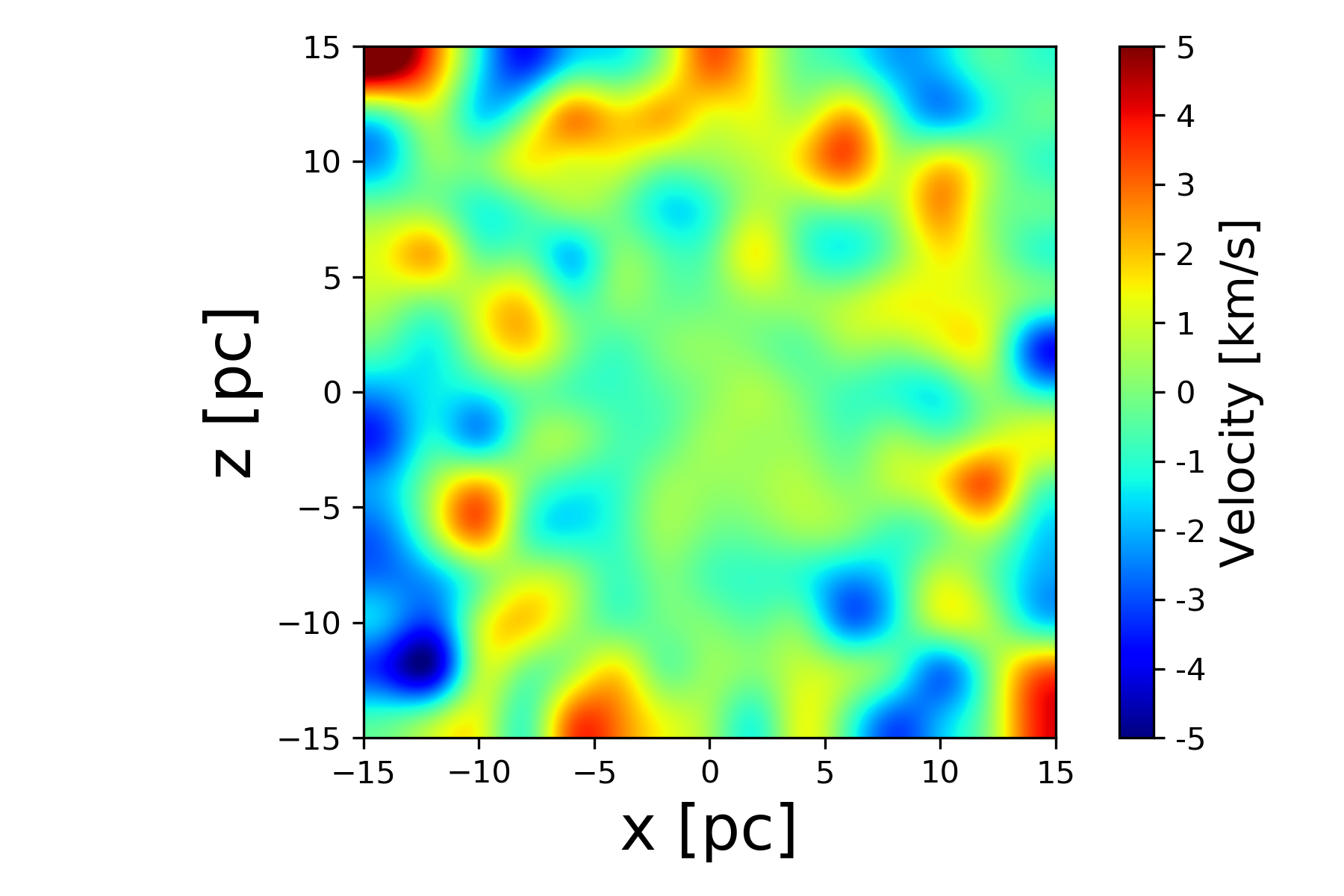}
	\includegraphics[trim={2.3cm 1 2.3 1},clip,width=0.45\textwidth]{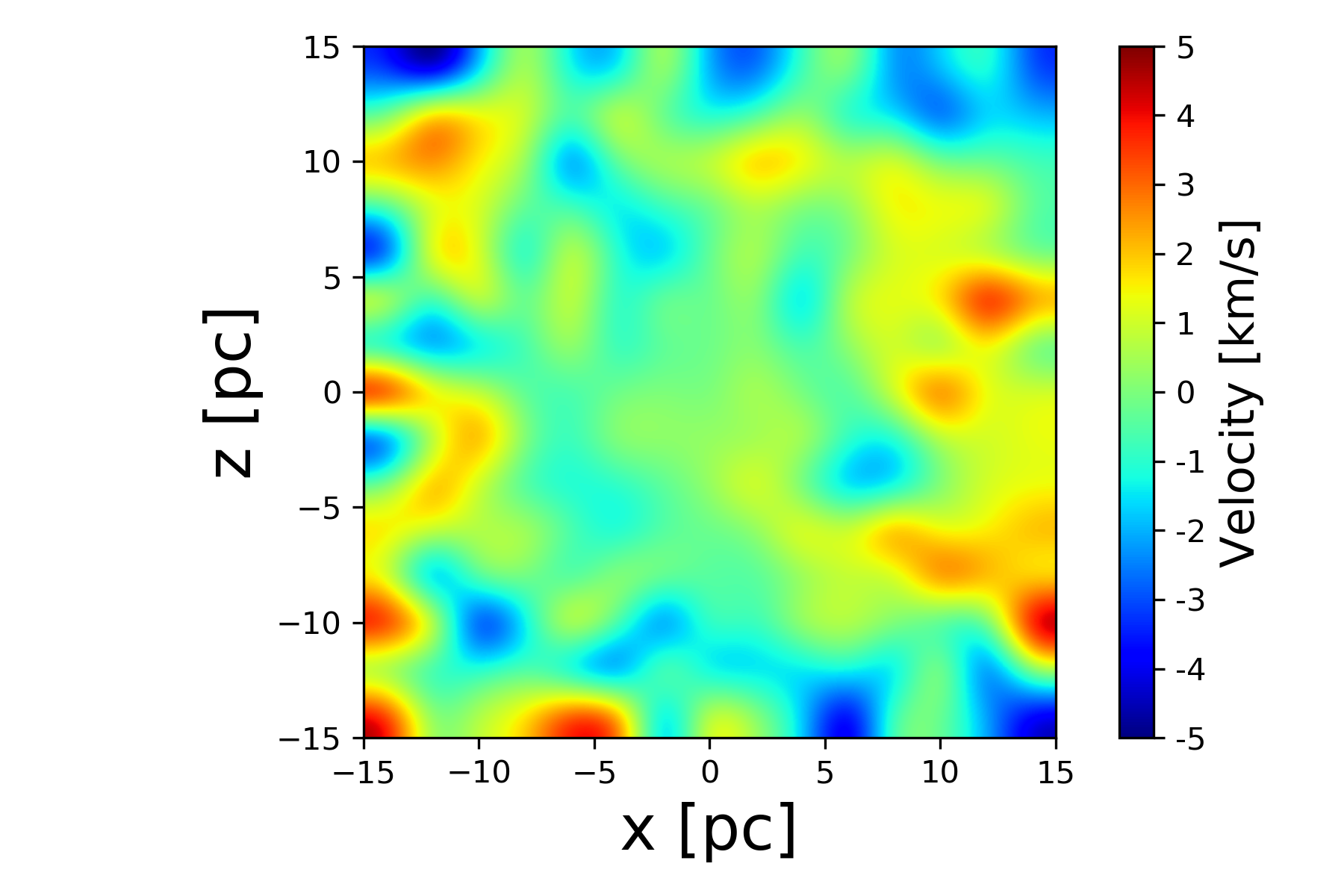}
	\includegraphics[trim={2.3cm 1 2.3 1},clip,width=0.45\textwidth]{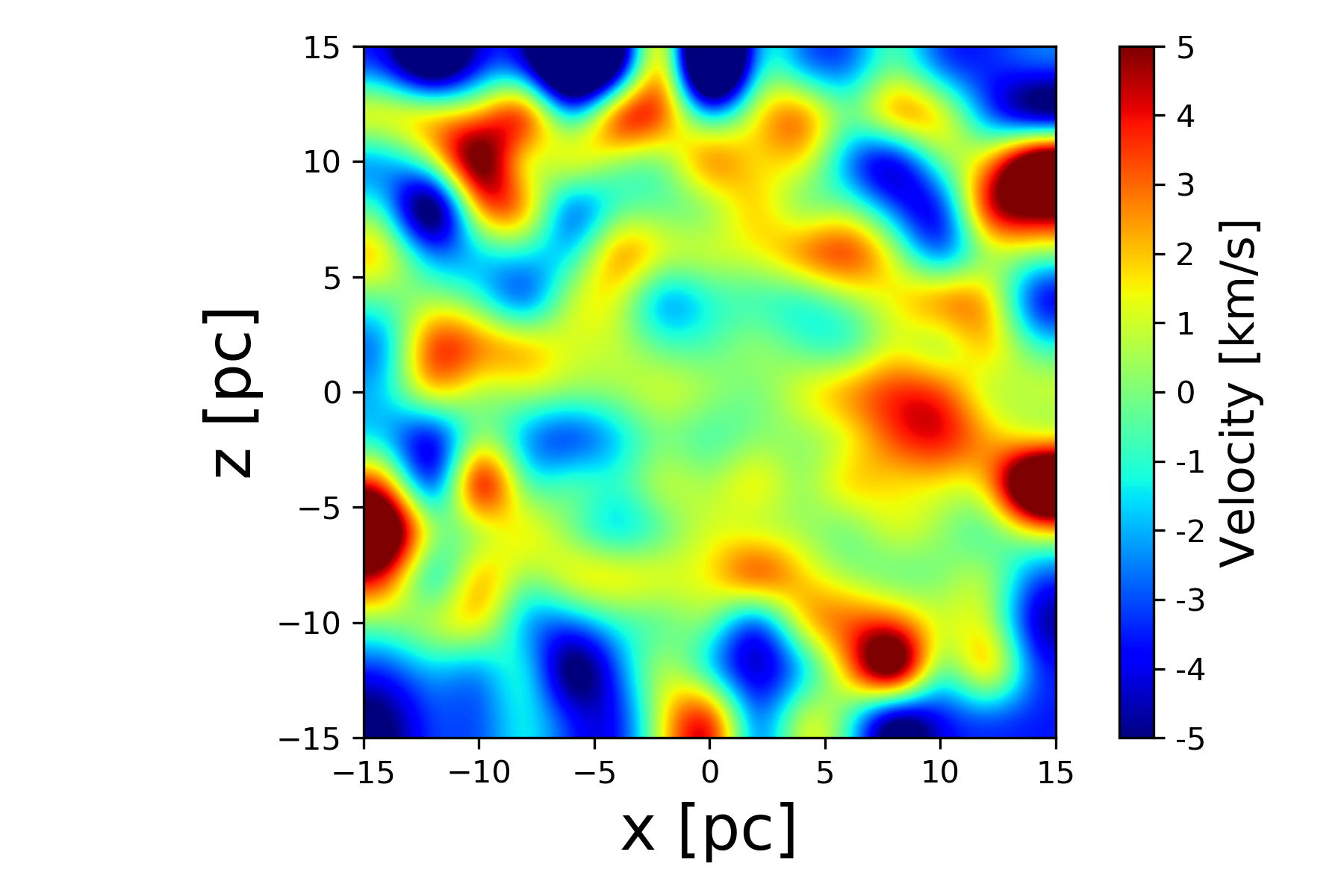}
	\includegraphics[trim={2.3cm 1 2.3 1},clip,width=0.45\textwidth]{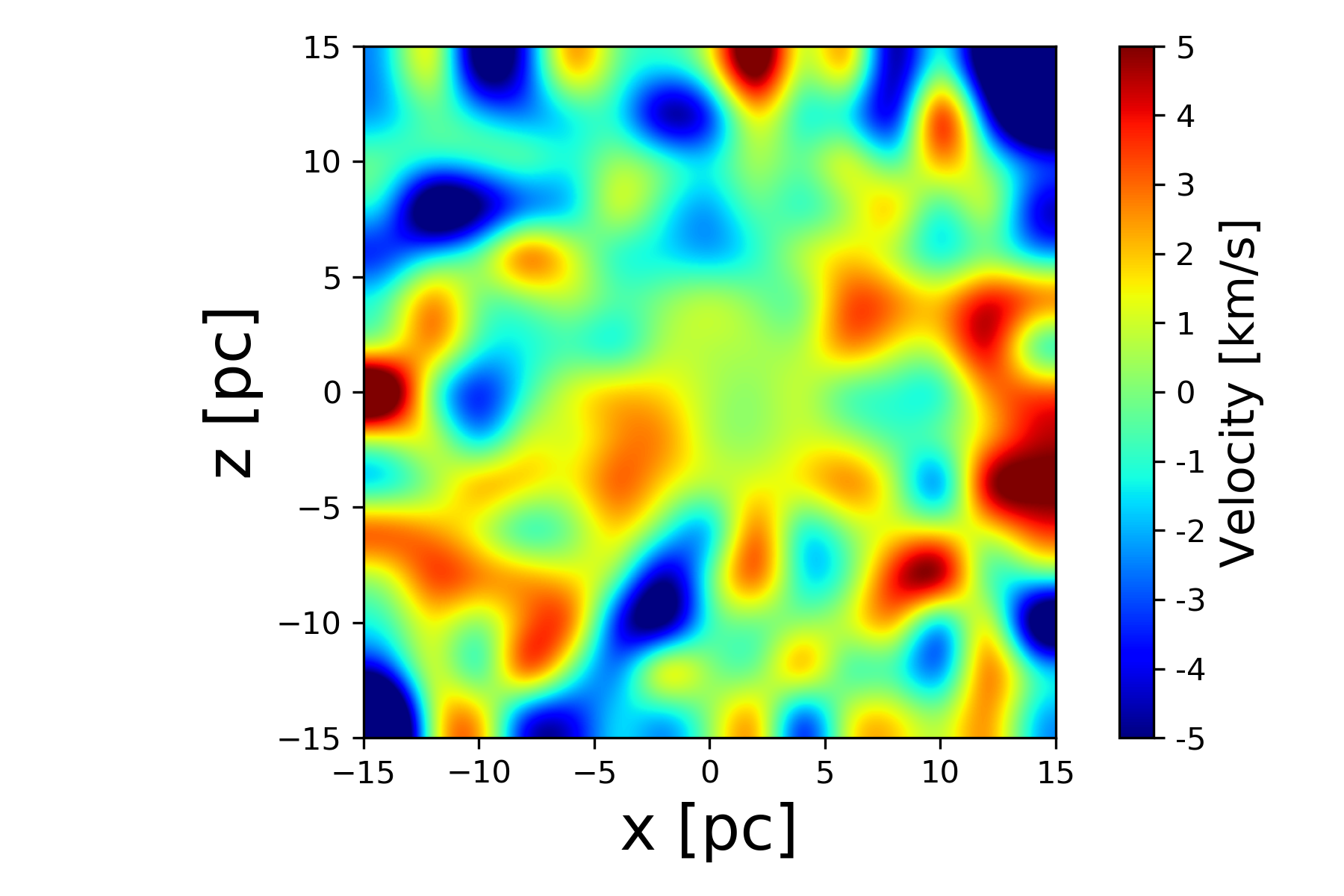}
	\includegraphics[trim={2.3cm 1 2.3 1},clip,width=0.45\textwidth]{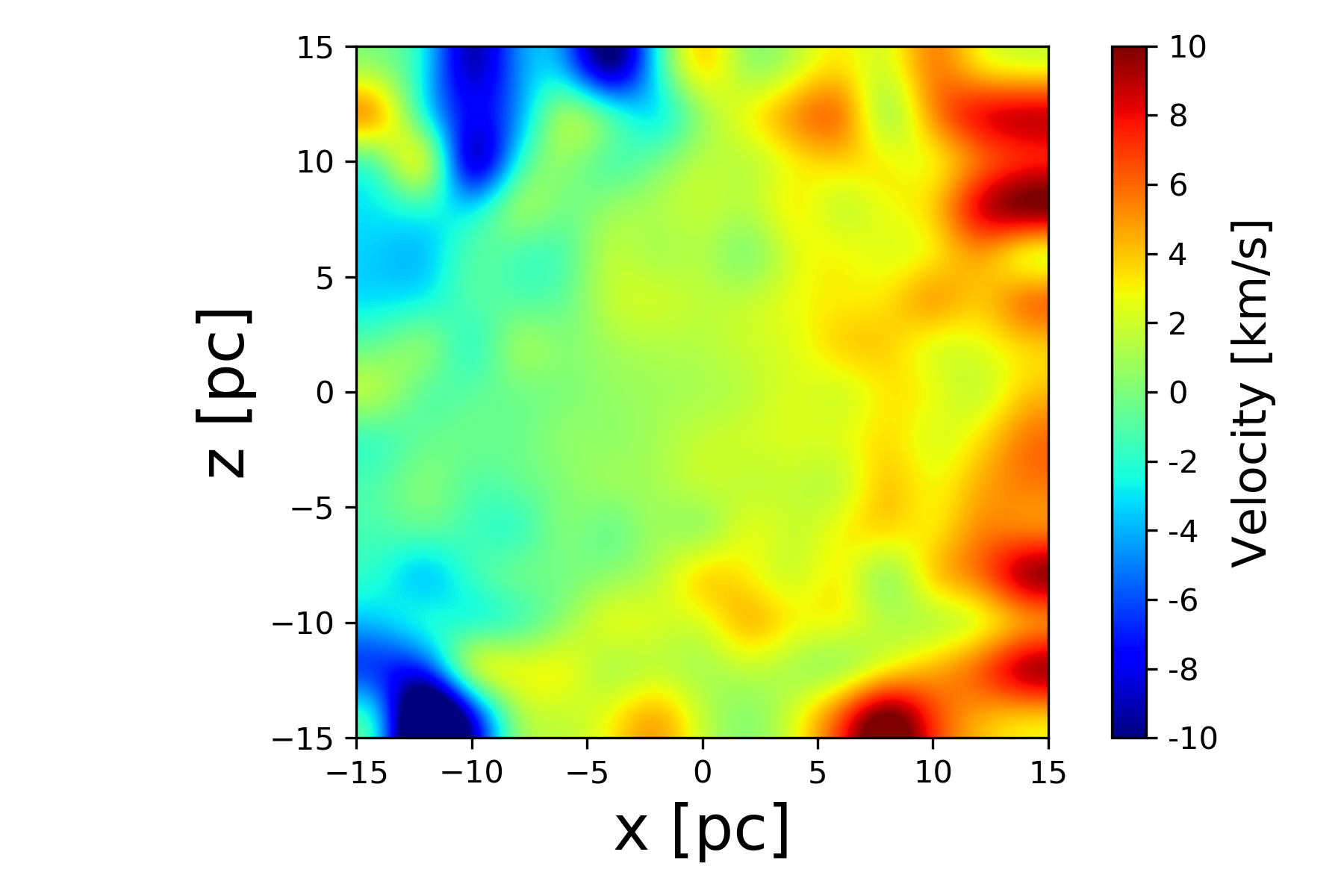}
	\includegraphics[trim={2.3cm 1 2.3 1},clip,width=0.45\textwidth]{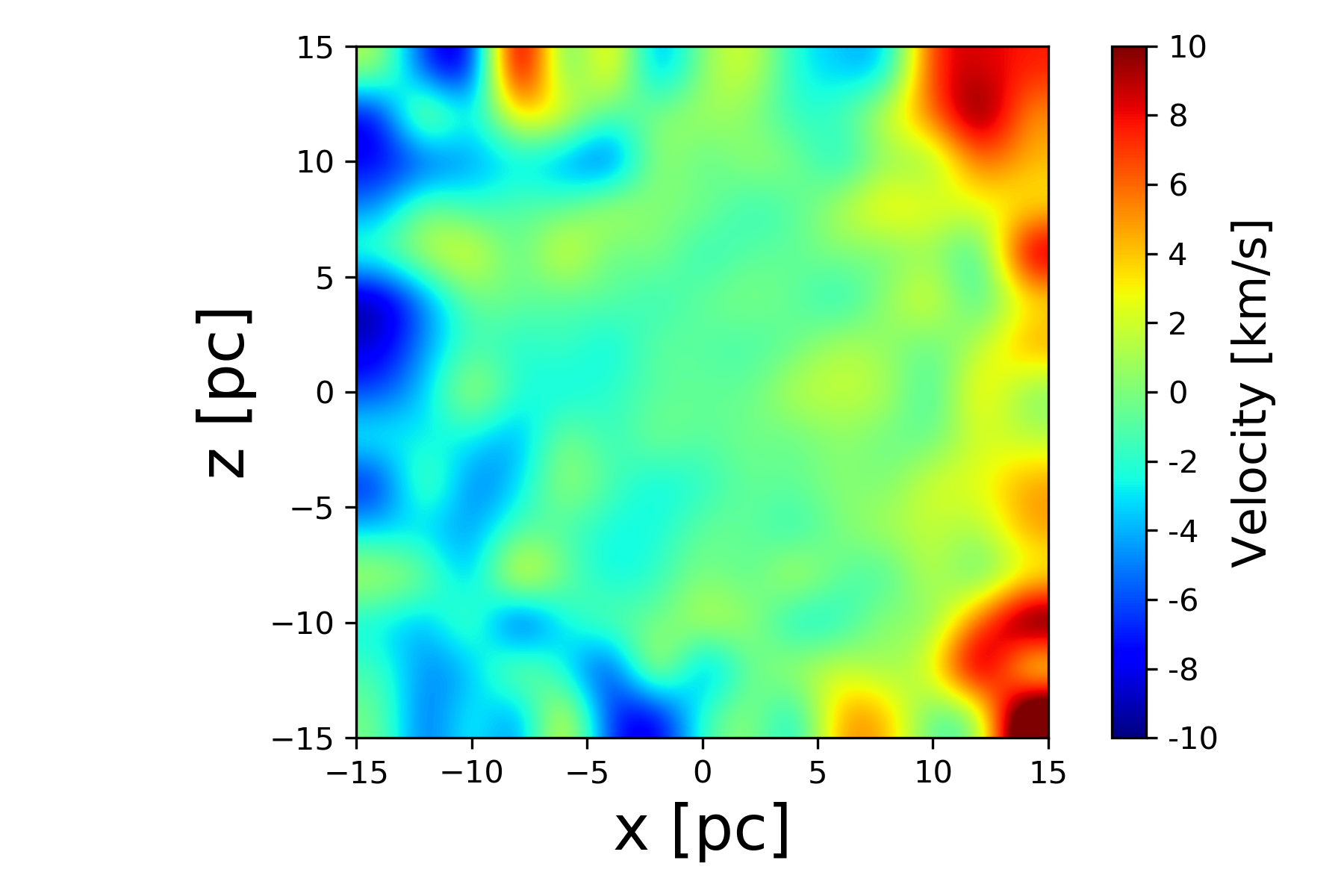}
    \caption{Velocity maps for the Core, Inter and Cusp simulations (from top to bottom, 1 displayed on the left and 2 on the right), illustrating the innermost $15\times15$\,pc of M54 at the end of each simulation. We find an increase in bulk movement with higher $\gamma$ values. A clear rotation of M54 cannot be seen for the Core and Inter simulations, as indicated by the slope in Figure \ref{fig:Velcurve}. For the Cusp cases, on the other hand, we find a clear rotation and bulk movement of M54.}
    \label{fig:VelmapC2W8}
\end{figure*}


\bsp	
\label{lastpage}
\end{document}